
\overfullrule=0pt
\parindent=24pt
\baselineskip=18pt
\magnification=1200

\input tables.tex


\def\simg{\vbox{\hbox{\raise0.90mm\hbox{$>$}}
\kern-18pt\hbox{\lower0.90mm\hbox{$\sim$}}}}
\def\siml{\vbox{\hbox{\raise0.90mm\hbox{$<$}}
\kern-18pt\hbox{\lower0.90mm\hbox{$\sim$}}}}

\def\alaph#1{\alpha_{#1}}
\def\alphagut{\alpha_G}
\def\at{A_t}
\def\bate#1{\beta_{#1}}
\def\bsgamma{b\rightarrow s\gamma}
\def\b#1{b_{#1}}
\def\e#1{E(#1)}
\def\f#1{F(#1)}
\def\m0{m_0}
\def\mew{M_{ew}}
\def\mgluino{m_{\tilde g}}
\def\mgut{M_G}
\def\mtop{m_t}
\def\mz{M_Z}
\def\tanbeta{\tan\beta}
\def\tildealpha#1{{{\tilde\alpha}_{#1}}}
\def\tildet#1{m^2_{{\tilde t}_{#1}}}
\def\vh#1{\langle H_{#1}\rangle}
\def\y#1{Y_{#1}}

{\nopagenumbers
\rightline{NUB-TH-3116/95}
\rightline{CTP-TAMU-07/95}
\rightline{February, 1995}
\vskip 10 pt
\centerline{\bf  Landau Pole Effects and the Parameter Space of
                 the Minimal Supergravity Model}
\medskip
\centerline{Pran Nath}
\centerline{Department of Physics, Northeastern University}
\centerline{Boston, Massachusetts 02115}
\medskip
\centerline{Jizhi Wu~~and~~Richard Arnowitt}
\centerline{Department of Physics, Texas A\&M University}
\centerline{College Station, Texas 77843}
\medskip
\medskip
\medskip
\centerline{\bf  Abstract}
\vskip 10pt
It is shown that analyses at the electroweak scale can be significantly
affected due to Landau pole effects in certain regions of the parameter space.
This phenomenon arises due to a large magnification of errors of the input
parameters $m_t$, $\alpha_G$ which have currently a 10 percent uncertainty in
their determination. The influence of the Landau pole on the constraint that
the scalar SUSY spectrum be  free of tachyons is also investigated.It is found
that this constraint is  very strong and eliminates a large portion of the
parameter space.Under the above constraint the trilinear soft SUSY breaking
term at the electroweak scale is found to lie in a restricted domain.
\vfill\eject}

\pageno=1

\noindent{\bf 1. Introduction}
\vskip 10pt

Currently among the models beyond the standard model, only those with
supergravity~[1] can accommodate the unification of the electroweak and strong
coupling constants that is consistent with the LEP data~[2],  provide a natural
mechanism to break supersymmetry via a hidden sector~[3] and can achieve a
radiative breaking of the electroweak symmetry breaking via renormalization
group effects[4]. The minimal supergravity model (MSGM) is one of the most
studied of such models~[4]. The parameter space of the MSGM with the radiative
symmetry breaking consists of only four parameters:
$$
\m0,~~\mgluino,~~\at,~~{\rm and}~~\tanbeta
\eqno(1.1)
$$
in addition to the top quark mass, $\mtop$(which is indicated by CDF to be
$\mtop=174\pm15$ GeV~[5]) and the GUT parameters, $\mgut$ (the grand
unification scale), and $\alphagut$ (the gauge coupling constant at $\mgut$).
Here in equation (1), $\m0$ is the universal scalar mass at $\mgut$, $\mgluino$
and $\at$ are the gluino mass and the trilinear scalar coupling at the
electroweak breaking scale, $\mew$ (we choose it to be the Z-boson mass,
$\mz=91.187$ GeV), and $\tanbeta=\vh2/\vh1$ is the Higgs vacuum expectation
value (VEV) ratio at $\mew$, where $\vh2$ and $\vh1$ give masses to the up and
down type quarks, respectively. The parameter space given by eq~(1) has two
branches, corresponding to the two signs of the supersymmetric Higgs mixing
parameter, $\mu$, since radiative symmetry breaking determines only the
magnitude of this parameter.

Although the MSGM has only four free parameters, in addition to the 19 for the
Standard Model[and 137 + 19 for the minimal supersymmetric standard model
(MSSM)], the parameter space is still quite large and one may put additional
constraints on the model to further reduce the allowed range of parameters.
Constraints that have been used in the past consist of proton stability for the
SU(5) type model~[6], the neutralino relic density constraint consistent with
the current estimates of dark matter in the universe~[7], the $\bsgamma$ decay
constraint~[8,9], as well as constraints on the lower bounds on the SUSY
particle masses from CDF, D0 and LEP experiments. Thus for models with proton
decay, such as the SU(5) type models, proton stability requires $\m0>\mgluino$
and $\tanbeta \siml 10$. The assumption of $R$ parity invariance implies that
the lightest neutralino (${\tilde Z}_1$) is the lightest supersymmetric
particle (LSP) for almost the entire parameter space. It does not decay due to
$R$ parity conservation, and hence may be regarded as the candidate for the
cold dark matter in the universe. Recent analyses have shown that there exists
a large domain in the parameter space which satisfies both the proton stability
and the relic density bounds~[7]. One may also impose additional constraints on
the theory such as the experimental constraint of the $\bsgamma$ branching
ratio. This decay is interesting as it is one of the few low-energy processes
that is sensitive to physics beyond the standard model~[10], and the first
experimental measurement of this process has been made~[11]. Recent analyses
[8,9] show that there is a significant region of  the parameter space where the
branching  ratio $B(\bsgamma)$ lies within the current experimental bound even
when combined with the relic density and proton stability constraints.

In addition to the constraints discussed in the previous paragraph, the
internal consistency of the theory also puts stringent constraints on the
parameter space. For example, the color $SU(3)_C$ group should remain unbroken
[12] including the one loop corrections to the effective potential
 [13] when the radiative symmetry breaking
is imposed, all scalar particles must be non-tachyonic, and the allowed
parameter space should be such that theory remains in the perturbative
domain.It turns out that the restriction that the mass spectrum be free of
tachyons imposes strong constraints on the parameter space of the model. An
interesting aspect of minimal supergravity unification is its predictivity.
There are 32 supersymmetric particles in the model whose massses are determined
in terms of the 4 parameters of eq. (1.1). Thus the theory makes 28
predictions. However, there is an important issue regarding how accurately
predictions of the mass spectra can be made. We shall show that the precision
with which predictions are made can be significantly influenced by the
proximity of the top quark to the fixed point in the top quark Yukawa coupling.
It turns out that in certain parts of the parameter space,the proximity to the
Landau pole can magnify errors in input data such as $m_t$ and $\alpha_G$
(which are known to no more than 10 percent accuracy) by an order of magnitude.
Thus in this region of the parameter space an accurate prediction of certain
subset of the SUSY mass spectra requires a knowledge of the input data to an
accuracy which is an order of magnitude more severe than what is currently
known. Further it is found that the proximity to the Landau pole often leads to
the lightest stop mass turning tachyonic which eliminates significant regions
of the parameter space.

The outline of this paper is as follows: In Sec.2, we give a brief review of
the Landau fixed point analysis for the top quark Yukawa coupling. We then
discuss two cases in which one can approach the Landau pole. These cases
correspond to fixing the input data (such as the value of the soft SUSY
breaking trilinear coupling) either at the GUT scale or at the electroweak
scale. In Sec.3, we discuss the magnification of errors that can occur when
one is close to the  Landau fixed point. The effect of these magnifications on
precision calculations at the electroweak scale is discussed. We also discuss
in this section the breakdown of perturbation theory as we approach the Landau
pole due appearance of $\log(\epsilon)/\epsilon$ terms (where $\epsilon$
measures the nearness to the Landau pole). In Sec. 4, we discuss the condition
that the scalar sector of the SUSY mass spectra not develop tachyonic modes. We
then use this contraint to determine the allowed range of $A_t$. Conclusions
are given in Sec. 5.

\vskip 10pt
\noindent{\bf 2. The Top Quark Landau Pole }
\vskip 10pt
Since the radiative breaking of the electroweak symmetry involves a large
extrapolation, from the grand unification scale $\mgut$ to the electroweak
scale $\mew$, the stability of solutions to small corrections needs to be
addressed. A number of possible corrections from various sources have been
discussed in the literature. These include two and higher loop corrections to
beta functions, threshold corrections at the GUT scale due to heavy particles
resulting from the breaking of the GUT group, threshold corrections at the
electroweak scale due to the light SUSY spectrum, loop corrections to the
effective potential and corrections due to possible higher dimensional
operators from quantum gravity effects. Some of these issues have been studied
in [14,15]. All of these corrections have influence on precision calculations
at the electroweak scale. In addition, there exist another type of corrections
that may be magnified due to proximity to the quasi fixed point of the top
quark
Yukawa Coupling~[16-20]. It is well known that the one-loop RG equation
analysis
using the gauge and Yukawa coupling evolution below $\mgut$ gives rise to a
quasi-infrared fixed point for the top quark Yukawa coupling corresponding to
$D_0=0$ where~[16]:
$$
    \y0={\y{t}\over\e{t} D_0},
\eqno(2.1)
$$
with
$$
  D_0=1-6\y{t}{\f{t}\over\e{t}}.
\eqno(2.2)
$$
Here in equations~(2.1) and~(2.2), $t=2\ln(\mgut/Q)$, $\y{t}=h_t^2/(4\pi)^2$,
$h_t$ is the top quark Yukawa coupling constant at the electroweak scale, $\y0$
is the value of $\y{t}$ at the GUT scale (i.e. at $t=0$) and Q is the running
mass scale. $\e{t}$ and $\f{t}$ are two form factors defined in [16] which only
depend on the gauge coupling RG equations:
$$
  \f{t}=\int_0^t \e{t} dt,
\eqno(2.3)
$$
and at the one loop level, $\e{t}$ maybe solved as,
$$
  \e{t}=(1+\bate3)^{16\over3\b3}(1+\bate2)^{3\over\b2}(1+\bate1)^{13\over9\b1},
\eqno(2.4)
$$
where $\bate{i}=\alaph{i}(0)\b{i}/4\pi$, $\b{i}$ are the coefficients of the
one loop beta functions, given by $(\b1,~\b2,~\b3)=(33/5,~1,~-3)$,
$\alaph{i}(0)$ are the gauge coupling constant at the GUT scale,
$\alaph{i}(0)=\alphagut$ and $\alaph1 = (5/3) \alaph{Y}$ with ${Y}$ being the
hypercharge in the Standard Model. From equation~(2.2), one observes that a
fixed point in the top quark Yukawa coupling corresponds to $\y{t}^f(t) =
\e{t}/ 6 \f{t}$. The existence of this fixed point implies a fixed point mass
for the top quark:
$$
\mtop^f=(8\pi/\alaph2(t))^{1/2}(\y{t}^f(t))^{1/2}\mz\cos\theta_w\sin\beta,
\eqno(2.5)
$$
where $\sin^2\theta_w=0.2328$ is the weak mixing angle. Thus,  for $\mgut =
10^{16.187} $~GeV, $\alphagut=1/24.0$, and $\mew=\mz$, one has $\e{t}=12.906$,
$\f{t}=263.954$, and hence $\y{t}^f(t) = 8.149 \times 10^{-3}$ and
$$
\mtop^f=2.163\mz\sin\beta=197.25\sin\beta.
\eqno(2.6)
$$
Because of the coupled nature of the RG equations, the same fixed point appears
in some of the other parameters of the theory. Near the fixed point small
variations in some of the input quantities can be magnified in the output.
Specifically, we shall show in the next section that the computed quantities in
certain  regions of the parameter space can be very sensitive to variations in
$\alphagut$, $\mtop$ and $\tanbeta$ near a fixed point.

As is evident from the discussion above all these sensitivities in the
parameters  can be traced back to the fact that in the parameter space one is
very close to the top quark Landau pole, which is determined by  $D_0$ being
close to zero. We note that $D_0$ depends on the gauge coupling constant RG
equations only through a combination $\e{t}/6\f{t}$ and on $\y{t}$ at the
electroweak scale. It is thus important to investigate the influence on this
quantity from variations of the input parameters such as the  grand unification
scale and the GUT coupling constants etc. We consider first the size of two
loop effects on the computation of the SUSY mass spectrum. In the following we
discuss the relative size of the two loop effects on $\alpha$'s relative to
variations in the value of $\alpha_G$. The two loop RG equations for the gauge
couplings are
$$
{d\tildealpha{i}\over dt}=
-[\b{i}+\sum_j{\b{ij}}\tildealpha{j}-a_{it}Y_t]\tildealpha{i}^2,
\eqno(2.7)
$$
where $\tildealpha{i}(t)=\alpha_{i}(t)/4\pi$, for $i,~j~=~1,~2,~3$.
In our analysis we have used the following values for
$\alpha_{i}$ at the electroweak scale,
$$
  \alaph1=0.016985\pm0.000020,\quad
  \alaph2=0.03358 \pm0.000011,  \quad
  \alaph3=0.118   \pm0.007.
\eqno(2.8)
$$
Table~1 displays a fit to the three gauge coupling constants using $\mgut =
10^{16.187} $~GeV and various values for $\alphagut$. One concludes that
$\alphagut=24.1$ gives the best fit to the numerical values for the three gauge
coupling constants given in equation~(2.8), both at the one loop level and at
the two loop level. One now infers from Table~1 that the variation in the GUT
coupling has much larger effects on the unification than the two loop RG
effects! We can similarly analyse  the two loop RG effects on the mass spectrum
and specifically on the top quark Landau pole. Table~2 shows a comparison
between the one loop and two loop evaluations for the form factors $\e{t}$,
$\f{t}$ and their ratio. It is seen in Table~2 that, although the one loop and
two loop calculations for $\e{t}$ and $\f{t}$ differ by several percent, their
ratio, and thus the fixed point for the top quark Yukawa coupling constant,
$\y{t}^f$, differ only by less than a percent. Again, we observe that the
changes in these quantities due to changes in $\alphagut$ are much larger than
the changes between the one loop and two loop calculations.

Next we discuss the question of what quantities may become singular as we
approach the Landau pole. This question depends critically on what is chosen as
the input and what as the output. For illustrative purposes we shall consider
two cases. In each of these cases we shall use $m_0$, $m_{1/2}$, $\tanbeta$,
$\alphagut$, $M_ G$, and $m_t$ as input. However, in Case I we shall assume
that the trilinear coupling $A_0$ is specified at the GUT scale, while in Case
II we shall assume that the trilinear coupling $A_t$ is specified at the
elecrtroweak scale. We discuss each of these cases in some detail. We consider
Case I first. Here the value of the trilinear coupling at the electroweak
scale, i.e., $A_t$ is given by
$$
A_t=A_0 D_0 + (m_{1/2}/m_0)(H_2-6 Y_t H_3/E)
\eqno(2.9)
$$
where the form factors, $H_2$, $H_3$ are defined in [16] and below. From the
above we see
that for fixed $A_0$ there is no pole in  $A_t$, rather one finds that $A_t$ is
smooth as one approaches the Landau pole, and in this limit we find that $A_t$
takes on the value
$$
A_{tP}=(m_{1/2}/m_0)(H_2-6 Y_t H_3/E)
\eqno(2.10)
$$
We see now that $A_{tP}$ is independent of the value of $A_0$ and is determined
in terms of other parameters. Thus there is a reduction by one in the number of
parametrers in this case. Further, here none of the computed quantities, such
as squark masses, etc., exhibit a pole. This case has been extensively
discussed in [18]. For Case II, $A_t$ is assumed fixed, and $A_0$ can be
computed in terms of  $A_t$ via the relation, $A_0=A_R/D_0$, so that
$$
 A_R=A_t-(m_{1/2}/m_0)(H_2-6H_3Y_t/E)
\eqno(2.11a)
$$
where
$$
 H_2=(\alpha_G /4 \pi) ( {16\over 3} h_3+3 h_2+{13\over 15} h_1)
\eqno(2.11b)
$$
and
$$
h_i=t(1+\beta_i t)^{-1}, \quad {H_3\over F}= t {E\over F} -1
\eqno(2.11c)
$$
Thus $A_0$ has a pole similar to the pole in the top Yukawa coupling constant.

We examine next the behavior of the Higgs parameter $\mu$ in the limit as one
approaches the Landau pole. We begin by noting that in the radiative breaking
of the electroweak symmetry, $\mu$ is determined by fixing the Z boson mass
$M_Z$ via the equation
$$
\mu^2={m^2_{H_1}-m^2_{H_2}\tan^2\beta \over \tan^2\beta-1}-{1\over2}\mz^2
\eqno(2.12)
$$
where
$$
  m^2_{H_1}=\m0^2+(0.3f_1+1.5f_2)\tildealpha{G}m_{1/2}^2
\eqno(2.13a)
$$
$$
 m^2_{H_2}=A_0 m_{1/2}f-(kA_0^2-h m_0^2)+em_{1/2}^2
\eqno(2.13b)
$$
Here $f_1$, $f_2$, $f$, $k$, $h$, and $e$ are defined in [16]. To investigate
the Landau pole limit we write the functions $h$, $k$, $f$, $e$ in the form
shown below using the identity $D_0~D=1$ where $D=1+6~Y_t~F(t)$. We have
$$
h=(3 D_0-1)/2
\eqno(2.14a)
$$
$$
k=3 Y_t F D_0/E
\eqno(2.14b)
$$
$$
f=-6 ~H_3 ~Y_t~D_0/E
\eqno(2.14c)
$$
$$
e=(3/2)[(G_1 D_0+Y_t G_2/E)+(H_2 D_0+6 H_4 Y_t/E)^2/3+H_8]
\eqno(2.14d)
$$
{}From the above we find that $m^2_{H_1}$ is smooth while $m^2_{H_2}$ develops
a
pole, i.e., one has
$$
  m^2_{H_2}(pole)=-3 Y_t (F/E) A_R^2/D_0
\eqno(2.15)
$$
{}From eq.~(2.12) one can obtain the limit of $\mu^2$ in the vicinity of the
Landau pole to be
$$
\mu^2=\mu_R^2/D_0+ \mu^2(NP)
\eqno(2.16a)
$$
where $\mu_R^2$ is given by
$$
\mu_R^2=3 {tan^2\beta \over tan^2\beta -1} Y_t{ F\over E} A_R^2 \eqno(2.16b)
$$
and where $\mu^2(NP)$ stands for the nonpole piece of $\mu^2$. We see from the
above that the limit of quantities depends strongly on whether $A_0$ or $A_t$
are assumed fixed as we approach the fixed point. For the case when $A_0$ is
kept fixed the limit is smooth, while the case when $A$ is kept fixed induces a
pole in several other parameters in the theory. The effect of these poles on
electroweak physics is investigated in the next section.

\vskip 10pt
\noindent{\bf 3.Effect of the Landau Pole on Physics at the Electroweak Scale}
\vskip 10pt

As discussed in Sec. II  for the case where $A_t$ is fixed, several of the soft
SUSY breaking parameters and the parameter $\mu^2$ develop Landau poles.
Computed quantities that depend on these parameters can also develop poles and
show rapid variations as we approach the fixed point. There are several
elements of the SUSY mass spectra which show such a behavior. We recall that
the stop quark mass matrix takes the following form,
$$
    \pmatrix{ \tildet{L} & \mtop(\at+\mu\cot\beta) \cr
                            \mtop(\at+\mu\cot\beta) & \tildet{R} \cr} .
\eqno(3.1)
$$
where
$$
\eqalignno{
\tildet{L}&={2\over3}\m0^2+{1\over3}m_{H_2}^2
                      +(-{1\over15}f_1+f_2+{8\over3}f_3)\tildealpha{G}m_{1/2}^2
                      +\mtop^2+({1\over2}-{2\over3}\sin^2\theta_w)\mz^2\cos2
\beta & \cr
& &(3.2)\cr
\tildet{R}&={1\over3}\m0^2+{2\over3}m_{H_2}^2
                      +({1\over3}f_1-f_2+{8\over3}f_3)\tildealpha{G}m_{1/2}^2
                      +\mtop^2+{2\over3}\sin^2\theta_w\mz^2\cos2\beta.&(3.3)\cr
}
$$
and $m^2_{H_1}$ and $m^2_{H_2}$ are as defined in eqs. (2.13). In the following
analysis we shall assume that $A_t$ at the electroweak scale is given and its
value at the grand unification scale $A_0$ is determined by eq.~(2.11).

The light stop mass is given by
$$
 \tildet1={1\over2}\left[(\tildet{L}+\tildet{R})
        -\sqrt{(\tildet{L}-\tildet{R})^2+4(\at+\mu\cot\beta)^2\mtop^2}\right].
\eqno(3.4)
 $$
while $\tildet2$ is given by a formula identical to eq.~(3.4) except that  the
relative sign between the two terms is positive. We can decompose $\tildet1$
and $\tildet2$ into a pole and an non-pole piece. We exhibit the pole part
below. For $\tildet1$ we have
$$
 \tildet1=-2x/D_0 +\tildet1(NP)
\eqno(3.5)
$$
where $x=Y_t A_R^2 F/E$.
 Similarly for $\tildet2$ we have
$$
\tildet2=-x/D_0 +\tildet2(NP)
\eqno(3.6)
$$
{}From the above it is seen that although $\tildet1$ is numerically smaller
than $\tildet2$ the pole part is larger in magnitude for  $\tildet1$ than for
$\tildet2$. Consequently the variation of $\tildet1$ is
always larger than the variation of $\tildet2$ as we
approach the quasi fixed point. Further the proximity of the fixed point makes
the
transition of $\tildet1$ to the tachyonic mode rapid and one expects large
regions of parameter space to be eliminated when one is near the fixed point.

We note that the first two generations of squarks and the three generations of
sleptons have in general a negligible correction due to the fact that
the $H_2$-Higgs coupling to their corresponding quark and lepton  masses is
negligible. Thus for these states one does not expect any rapid variations in
the vicinity of the  Landau pole. However, rapid variations  occur for some of
the chargino and the  neutralino spectrum as we approach the Landau pole. This
can be seen most easily in the scaling limit where the chargino spectrum is
given by
$$
m_{\tilde W_1}\simeq\tilde m_2,\quad  m_{\tilde W_2}\simeq \mu
\eqno(3.7)
$$
where $\tilde m_2=(\alpha_2/\alpha_G)m_{1/2}$. From the above we infer that
$m_{\tilde W_1}$ will show no variation while $m_{\tilde W_2}$ will show a
significant variation as we approach the Landau pole. Similarly the neutralino
masses in the scaling limit are given by
$$
m_{\tilde Z_1} \simeq m_{\tilde W_1}/2,\quad m_{\tilde Z_2} \simeq m_{\tilde
W_1} \eqno(3.8a)
$$
$$
 m_{\tilde Z_3} \simeq \mu, \quad m_{\tilde Z_4} \simeq  \mu
\eqno(3.8b)
$$
Here again as we approach the fixed point we do not expect any rapid variation
in $m_{\tilde Z_1}$ and $m_{\tilde Z_2}$. However, $m_{\tilde Z_3}$ and
$m_{\tilde Z_4}$ are expected to show a rapid variation in the vicinity of the
Landau pole. These expectations are borne out in numerical analyses of the
chargino and neutralino mass spectra. A similar analysis can be carried out for
the Higgs. For the lightest Higgs the parameter $\mu$ enters only at the loop
level and its dependence in the lightest Higgs mass is suppressed. Thus no
rapid variation is expected for the lightest Higgs in the vicinity of the
Landau pole. The three remaining Higgs in the scaling limit are all
approximately degenerate with a common mass $m_A^2 \simeq 2
m_3^2/sin2\beta$.One can determine $m_3^2$  from the radiative breaking
equation
$$
sin2\beta = 2 m_3^2/(2\mu^2+m_{H_1}^2+m_{H_2}^2)
\eqno(3.9a)
$$
Using Eqs(2.15) and (2.16) one finds
$$
m_A^2=-{3\over {cos(2\beta)}} Y_t  {F\over E}A_R^2 {1\over D_0}+m_A^2(NP)
\eqno(3.9b)
$$
We see that the three heavy Higgs possess a pole piece  and thus show a rapid
variation as we approach the Landau pole. These expectations are again borne
out by numerical computations of the mass spectra. $m_3^2$ is determined by the
 renormalization group to be
$$
 m_3^2 = -B_0\mu^2+r\mu_0m_{1/2}+s A_0\mu_0
\eqno(3.10)
$$
Here $\mu^2=~\mu_0 ^2~ q$ and $r$, $s$, $q$ are defined in [16].One finds then
that $B_0$ becomes singular as we approach the Landau pole.

Next we discuss the effect on electroweak physics due to small changes in the
input parameters when one is in the vicinity of the Landau pole. The input
parameters for the case under discussion are $p_i = (m_0, ~m_{1/2}, ~A_t,
{}~\tanbeta, ~\alpha_G, ~M_G, ~m_t)$, $ i=1,...,7$. Let us denote by $Q_a$ the
output quantities which exhibit a pole structure near the fixed point:
$$
Q_a=C_a/D_0 +Q_a(NP)
\eqno(3.11)
$$
We denote by $Q_{a,i}$ the derivative of $Q_a$ with respect to $p_i$. Then
$$
Q_{a,i}=C_{a,i}/D_0-C_a~D_{0,i}/D_0^2+Q_a(NP)_{,i}
\eqno(3.12)
$$
It is now seen that the largest variations arise for those parameters for which
$D_{0,i}$ are non-vanishing. There are three such parameters: $m_t$,
$\alpha_G$, and $\tanbeta$. We shall call these order two parameters since
their variations involve a double pole. The effect on electroweak physics of
the variations of these parameters is important since there are currently
significant experimental errors in the determinations of, for example, $m_t$,
and $\alpha_G$. These experimental errors can be vastly exaggerated if one is
in the vicinity of the Landau pole. We exhibit this maginification of errors in
Fig1. Here the $\tildet1$ mass is plotted as a function of $m_t$ for various
assumed values of  $\tanbeta$. In the example shown (for the case $A_t=0.5$)
one finds that the $\tildet1$ mass begins to show a very rapid variations as we
approach the Landau pole. Thus in the vicinity of the Landau pole changes in
$m_t$ of the order of a few GeV can lead to changes in the $\tildet1$ mass of
several hundred GeV. Similarly a small fractional change in $\tanbeta$ can
generate a similar large change in $\tildet1$ mass. A similar phenomenon occurs
for variations in $\alpha_G$. This means that precision analyses of mass
spectra will require a very high degree of accuracy in the parameters $m_t$,
$\alpha_G$ if one happens to lie in the vicinity of the Landau pole.

Fig 1 also
shows that the stop 1 turns tachyonic as we approach the Landau pole.
It can be seen from eq(3.5) that the residue of the pole term
is negative  and thus as $m_t$ approaches $m_t^f$ the pole term cancels the
non-pole term in eq(3.5) turning $\tilde t_1$ tachyonic.We wish to point  out
the important role that the trilinear coupling $A_t$ and its sign play in
turning $\tilde t_1$ tachyonic. Actually the parameter  which controls the
approach to tachyonic limit is $A_R$ which is defined by eq(2.11). In the
vicinity of the pole $x\simeq A_R^2/6$. To
get an idea of the sizes of various entries in eq(2.11a) we find that for the
same parameters as used in eq(2.6) one has
$$
A_R\simeq A_t- 0.613 m_{\tilde g} \eqno(3.13)
$$
One finds then that for $A_t$ positive there is a cancellation between the two
terms on the right hand side of eq(3.13) which reduces the size of $A_R$ and
slows down the approach to the tachyonic limit.Thus  for positive $A_t$ one
will have to get relatively close to the Landau pole for $\tilde t_1$ to turn
tachyonic.For the $A_t$ negative case there is  a reinforcement between the two
terms on the right hand side of eq(3.13). Thus in this case approach to the
tachyonic limit is much faster. An illustration of these results is given in
Fig 2. Here one sees  clearly  the strong influence that the Landau pole has in
turning the stop 1 tachyonic. A more detailed investigation of the tachyonic
condition is given in Sec 4,where the tachyonic limit is used to delineate the
allowed parameter space of MSGM.

 The coefficient of the double pole term exhibits certain scaling laws. Thus
neglecting the single pole and the non-pole terms one has
$$
Q_{a,i/}Q_{a,j} = ~D_{0,i}/D_{0,j}
\eqno(3.14)
$$
{}From the above we see that the ratio of variations is independent of the
index
$a$ and hence also independent of all the parameters that do not enter in
$D_0$. If we label the order two variables $m_t$, $\alpha_G$, and $sin\beta$ by
$i=1,2,3$, then
$$
Q_{a,1}/Q_{a,3}= -sin\beta/m_t
\eqno(3.15a)
$$
$$
Q_{a,2}/Q_{a,1}= -(1/2) m_t~d(\log(F/E))/d(\alpha_G)
\eqno(3.15b)
$$
We emphasize, however, that there are important non pole corrections which one
must include in any realistic evaluation of the variations.

We wish to investigate next the issue of how close one may approach the Landau
pole. First from the stability of the the potential at the GUT scale one
requires that
$$
A_0^2< 3(3m_0^2+\mu_0^2)
\eqno(3.16)
$$
Now
$$
\mu^2=\mu_0^2 (1+\beta_2 t)^{3/b_2}(1+\beta_1 t)^{3/b_1} (D_0)^{1/2}
\eqno(3.17)
$$
We have already seen that from radiative breaking of the electro-weak symmetry
one deduces $\mu^2$ to have the form $\mu^2=a_1+a_2/D_0$. Using these results
we can deduce the $D_0$ dependence of $\mu_0^2$. One finds
$$
\mu_0^2=a/D_0^{1/2}+b/D_0^{3/2}
\eqno(3.18)
$$
Using eq.~(3.18) the condition of stability of the potential at the GUT scale
takes the form
$$
3m_0^2 D_0^2+ a D_0^{3/2}+bD_0^{1/2}-A_R^2/3>0
\eqno(3.19)
$$
If one is close enough to the pole that powers of $D_0$ higher than $D_0^{1/2}$
can be neglected then one obtains
$$
D_0^{1/2}>A_R^2/(3b)
\eqno(3.20)
$$

The above equation gives the point of closest approach to the Landau pole.
However, as one gets very close to the fixed point other phenomenon must be
taken into account. These include higher order terms in the top quark Yukawa
coupling equations, loop effects in the radiative breaking equations etc. Here
we point out an interesting phenomenon associated with inclusion of loop
effects in radiative breaking and its implication in the Landau pole analysis.
With inclusion of one-loop effects to the effective potential, the condition of
eq.~(2.12) is modified as follows:
$$
\mu^2 = { m^2_{H_1} + \Sigma^1 - (m^2_{H_2}+\Sigma^2) \tan^2 \beta \over \tan^2
\beta-1}
- {1\over2} \mz^2
\eqno(3.21)
$$
where $\Sigma^{1,2}$ are the one loop corrections. The largest contributions to
$\Sigma^{1,2}$ arise from the stops and are given in the Appendix. Using
results of the Appendix in eq.~(3.21) we find that the satisfaction of
radiative electro-weak symmetry breaking now shows that the solution of $\mu^2$
 with the above corrections is more complicated. It has the form
$$
\mu^2=a_1+b_1/\epsilon+c_1log(d_1 \epsilon)/\epsilon
\eqno(3.22)
$$
where $\epsilon=D_0$. From eq.~(3.22), we see that inclusion of the one loop
correction in the effective potential brings in the new feature of the
$log(\epsilon) /\epsilon$ term while the tree contribution is only
O($1/\epsilon$). This means that as we approach the Landau pole the
contribution of the one loop effective potential eventually becomes larger than
the tree contribution due to the $log(\epsilon)$ divergence. This would signal
the breakdown of the perturbation theory. Thus perturbation theory puts a
natural cutoff on how close one can approach the Landau pole. The point of
closest approach is defined by the condition that the loop contribution in
eq.~(3.22) be smaller than the tree contribution.

\vskip 10pt
\noindent{\bf 4. The Parameter Space of the Minimal Supergravity Model}
\vskip 10pt
In this section we shall discuss the implications of the condition that the
SUSY spectrum be free of tachyons in the analysis of the radiative electroweak
symmetry breaking. As it turns out, of the 32 SUSY particles, only the light
stop may become tachyonic in some region of the parameters. From eq.~(3.5) it
is easy to understand why the light stop can become tachyonic.
 Since $x>0$, the first term of Eq. (3.5) will always
dominate the second for $D_0$ small enough. Thus as one approaches the Landau
pole, $\tildet1$ will always turn tachyonic, provided eq. (3.20) is not
violated. One can then ask when the light stop becomes tachyonic. The answer is
very sensitive to the values one chooses for the parameters, since $c$, $e$,
$f$, $f_1$, $f_2$, $f_3$, $h$, and $k$ have very complicated dependence on
$\mgut$, $\alphagut$, $\mtop$ (via $\y{t}$), and $\tanbeta$. To determine if
the light stop is tachyonic, we need to solve the equation $\tildet1=0$.
Fortunately, this equation can be solved analytically for $\at=0$. We consider
two points in the parameter space for $\at=0$, $\tanbeta=1.732$, and
$\mtop=150$~GeV.

Case 1):We choose the remaining parameters to be $\mgut=10^{16.1}$~GeV,
$\alphagut=1/25.7$ (we may replace
$\mz^2\cos^2\theta_W$ by $M_W^2$ while computing $\y{t}$ at the electroweak
scale), and $\alaph2=0.03322$. Scaling all the masses by 100 GeV, the light
stop has zero mass on the $m_{1/2}$ and $\m0$ plane along the following line:
$$
  19.66-15.11m^2_{1/2}-4.54m^4_{1/2}+3.77\m0^2
  +0.74m^2_{1/2}\m0^2+0.39\m0^4=0,
\eqno(4.1)
$$

Case 2): We choose remaining  parameters to be $\mgut=10^{16.187}$~GeV,
$\alphagut=1/24.11$,
and $\alaph2=0.03358$. Using the same scaling as in case 1, the light stop has
zero mass on the $m_{1/2}$ and $\m0$ plane along the following line:
$$
  19.04+28.44m^2_{1/2}+18.41m^4_{1/2}+6.16\m0^2
  +6.47m^2_{1/2}\m0^2+0.55\m0^4=0,
\eqno(4.2)
$$

There is  a significant difference between Case 1 and Case 2. For Case 1,
coefficients in eq~(4.1) appear with both positive and negative signs, while
in Case 2, coefficients in eq~(4.2) all have positive signs. Thus, the
condition that the light stop be tachyonic may be satisfied in Case 1 for
points below the line given by equation~(4.1), while this same condition cannot
be satisfied at any point on the $\m0$-$m_{1/2}$ plane for Case 2, even though
the two cases differ by only a small change in $\alphagut$ and other
parameters. Cases 1 and 2 discussed above were for illustrative purposes only.
We discuss next a more systematic study of the condition that the light stop be
nontachyonic. Figures 3~and~4 show the results for $\mtop=170$ GeV
(corresponding to a pole mass of about 175 GeV), and $\tanbeta=5.0$. Fig.~3
shows the results for $\at/m_0=-0.25$, $-0.3$, $-0.4$, $-0.5$, and $-0.8$ on
the $\mgluino$-$\m0$ plane, the light stop is tachyonic above and right to each
curve for a fixed $\at$. Fig.~4 demonstrates the results for $\at/m_0=0.5$,
$0.8$, $1.0$, $2.0$, and $5.0$ on the $\mgluino$-$\m0$ plane, the light stop is
tachyonic above and left to each curve for a fixed $\at$. From these figures,
one observes that the condition that the light stop be nontachyonic excludes
more parameter space when $\mu$ and $\at$ have the same sign than when they
have opposite signs. Further large values of $\at$ of either sign, specifically
values of $\at<-0.8 m_0$ and $\at>5.0 m_0$, are excluded by this condition.
Thus the  allowed region of $A_t$ lies within  $-0.8 m_0<\at<5.0 m_0$ for
$tan\beta$=5 and $m_t$=170 GeV. This is
a very strong constraint and it is interesting that the light stop being
tachyonic alone can impose such a stringent constraint on the parameter space
of MSGM.

\vskip 10pt
\noindent{\bf 5. Conclusions}
\vskip 10pt
We have investigated in this paper the phenomena associated with effects of the
Landau pole in the top quark Yukawa coupling and implications of the constraint
that the SUSY spectrum be free of tachyons. Regarding the first we find that in
the renormalization group analyses some of the mass spectra (i.e., masses of
the stops, the heavy chargino, heavy neutralinos, and the heavy Higgs) are very
sensitive to the input errors in $m_t$, $\alpha_G$, and $\tanbeta$ when the top
mass is close to its fixed point value. In this region of the parameter space
precision physics requires much greater accuracy of the input data. We also
investigated the issue of how close one may approach the fixed point. Regarding
the second phenomenon the condition that the SUSY spectrum be tachyon free
turns out to be a rather stringent one as it eliminates a large part of the
parameter space. For example, the trilinear scalar coupling, i.e., $\at$ is
confined within the range of $-0.8 m_0$ to $5.0 m_0$, for the parameters of
Fig. 3 and 4. The graphs indicate the strong correlation between $\at$ and
$\mu$, i.e., the light stop is more likely to become tachyonic when $\at$ and
$\mu$ have the same sign.
\vskip 10pt
\noindent{\bf  Acknowledgements}
\vskip 10pt
This research was supported in part by NSF grant numbers PHY-19306906 and
PHY-9411543.
\vskip 10pt
\noindent{\bf  Appendix}

The contribution to $\Sigma^1$ from stops ${\tilde t}_{1,2}$ is given by
$$
\eqalignno{
\Sigma^2_{{\tilde t}_{1,2}} =& 3 {{\alpha_2 + \alpha_Y} \over 8\pi}
{\tildet{1,2}} [ { 1\over4} \mp {1\over ({\tildet2}  - {\tildet1})}
\{ {1\over2} (\tildet{L} - \tildet{R}) ({1\over2} - {4\over3} \sin^2\theta_W)
\cr &~~~+ {m^2_t \over M^2_Z \sin^2\beta} \mu (\mu  +tan\beta
A_t)\}]\log ( { {\tildet{1,2}} \over e Q^2 } ). &(A.1)\cr}
$$

Similarly the contribution to $\Sigma^2$ from stops ${\tilde t}_{1,2}$ is
$$
\eqalignno{
\Sigma^2_{{\tilde t}_{1,2}} =& 3{{\alpha_2 + \alpha_Y} \over 8\pi}
{\tildet{1,2}} [ {- 1\over4} + {m^2_t \over M^2_Z \sin^2\beta} \mp {1\over
({\tildet2}  - {\tildet1})} \{ {1\over2} (\tildet{L} - \tildet{R}) (-{1\over2}
+ {4\over3} \sin^2\theta_W) \cr &~~~+ {m^2_t \over M^2_Z \sin^2\beta} m_0 A_t
(\mu \cot\beta + m_0 A_t)\}]\log ( { {\tildet{1,2}} \over e Q^2 } ).
&(A.2)\cr}
$$

\vskip 10pt
\noindent{\bf  References}
\vskip 10pt

\item{[1]} For reviews, see: P. Nath, R. Arnowitt, and A.H. Chamseddine,
"Applied N=1 Supergravity", (World Scientific, Singapore, 1984); H.P. Nilles,
Phys. Rep. 110, 1 (1984).

\item{[2]} P. Langacker, Proceedings, PASCOS 90 Symposium, Editors, P. Nath and
S. Reucroft (World Scientific, Singapore, 1990); J. Ellis, S. Kelly, and D.V.
Nanopoulos, Phys. Lett. {\bf 249B}, 441 (1990), {\it ibid}, {\bf 260B},
131(1991); U. Amaldi, W. de Boer, and H. Furstenau, {\it ibid} {\bf 260B}, 447
(1991); P.Langacker and N. Polonsky, Phys. Rev. {\bf D47}, 4028 (1993).

\item{[3]} J. Polonyi, Univ. of Budapest Rep. No. KFKI-1977-93 (1977); A. H.
Chamseddine, R. Arnowitt, and P. Nath, Phys. Rev. Lett. {\bf 29}, 970 (1982).

\item{[4]}For a recent review, see: R. Arnowitt, and P. Nath, Lectures at VII
J.A. Swieca Summer School, Campos de Jordao, Brazil, 1993 (World Scientific,
Singapore, 1994).

\item{[5]} CDF Collaboration, Fermilab-Pub-94/097-E (1994).

\item{[6]} P. Nath, A. H. Chamseddine, and R. Arnowitt, Phys. Rev.{\bf D32},
2348 (1985); R. Arnowitt, and P. Nath, Phys. Rev. Lett. {\bf 69}, 725 (1992),
Phys. Lett. {\bf 289B}, 368 (1992);J.~Hisano,H.~Murayama,and T.~Yanagida,
Nucl.Phys.{\bf B402}, 406(1993).

\item{[7]} R. Arnowitt, and P. Nath, Phys. Lett. {\bf 299B}, 58 (1993) and
Erratum {\it ibid} {\bf 303B}, 403 (1993); P. Nath, and R. Arnowitt, Phys. Rev.
Lett. {\bf 70} 3696 (1993) ; J. Lopez, D. Nanopoulos, and K. Yuan, Phys. Rev.
{\bf D48},2766(1993) ;M.~Drees,and M.~M.~Nojiri,Phys.Rev.{\bf D48},3483(1993);
 A.Bottino,V.de Alfero,N.~Forengo,G.~Mignola,and M.\par
Pignone,Astro. Phys. {\bf 2},67(1992).

\item{[8]}  J. Wu, R. Arnowitt, and P. Nath, to appear in  Phys. Rev.{\bf
D51} (1995).

\item{[9]} P. Nath, and R. Arnowitt, Phys. Lett. B336, 395 (1994).

\item{[10]} S. Bertolini, F. Borzumati, A. Masiero, and G. Ridolfi, Nucl. Phys.
{\bf B353}, 541 (1983); J. Hewett, Phys. Rev. Lett. {\bf 70}, 1045 (1993); V.
Barger, M. Berger, and R.J.N. Phillips, Phys. Rev. Lett. {\bf 70}, 1368 (1993);
M.A. Diaz, Phys. Lett. {\bf 304B}, 278 (1993); R. Barbieri, and G. Guidice,
Phys. Lett. {\bf 309B}, 86 (1993); J.L. Lopez, D.V. Nanopoulos, and G. Park,
Phys. Rev.{\bf D48}, R974 (1993); Y. Okada, Phys. Lett. {\bf 315B}, 119
(1993); R. Garisto, and J.N. Ng, Phys. Lett. {\bf 315B}, 372 (1993)); G. Kane,
 C. Kolda, L.Roszkowski, and D.J. Wells, Phys. Rev. {\bf D49}, 6173 (1994);
 S.Bertolini, and F. Vissani, SISSA 40/94/EP (1994).

\item{[11]} E.H. Thorndike, Talk at ICHEP94, Glasgow, 1994 (ICHEP94 Ref.
GLS0392); R. Ammar {\it et. al.}(CLEO Collaboration), Phys. Rev. Lett. {\bf
71},
674 (1993).

\item{[12]} J.M. Frere, D. Jones, and S. Raby, Nucl. Phys. {\bf B222}, 1
(1983); M. Claudson, J. Hall, and I. Hinchliffe, Nucl. Phys. {\bf B288}, 501
(1983); M. Drees, M. Gluck, and K. Brassie, Phys. Lett. {\bf 157B}, 164 (1985).

\item{[13]}For example see, R. Arnowitt, and P. Nath, Phys. Rev.{\bf D46},
3981 (1992), and the references quoted therein.

\item{[14]} D. Ring, S. Urano, and R. Arnowitt, Texas A\&M Report,
CTP-TAMU-46/94 (1994).

\item{[15]} T. Dasgupta, P. Mamales, and P. Nath, NUB-TH-3115/94 (1994).

\item{[16]} Iba\~nez, C. Lopez, and C. Mu\~nos, Nucl. Phys. {\bf B256}, 218
(1985).
\item{[17]} C.Hill, Phys.Rev.{\bf D 24},691(1981).
\item{[18]} V. Barger,M.S.Berger and P.Ohman,Phys.Lett.{\bf B314},351(1993).

\item{[19]} W. A. Bardeen, M. Carena, S. Pokorski, and C. E. M. Wagner,
Phys.Lett. {\bf B20}, 110 (1994);  M. Carena, and C.E.M. Wagner,
CERN-TH.7393/94 (1994).

\item{[20]} M. Carena, M. Olechowski, S. Pokorski, and C.E.M. Wagner, Nucl.
Phys. {\bf B419}, 213 (1994).

\vskip 10pt
\noindent{\bf Table Captions}
\vskip 10pt

\noindent{\bf Table 1} The fit to the electroweak data for the gauge coupling
constants given in equation~(2.9) in one and two loop approximations, for
$\mgut = 10^{16.187}$ GeV, and various values for $\alphagut$.

\noindent{\bf Table 2} The effects of the variation of $\alpha_{GUT}$ on the
ratio F/E and hence on the top Landau pole in one and two loop R.G.analysis.
One finds that the variations of $\alpha_{GUT}$ are much more significant
than the two loops effects.

\vskip 10pt
\noindent{\bf Figure Captions}

\noindent{\bf Fig. 1} Exhibition of $\tildet1$ vs.$m_t$ for the case when
$A_t=0.5 m_0$, $m_0=600$ GeV, $\mgluino=300$ GeV and
$\tanbeta=1.2,1.4,2.0,3.0,5.0,7.0,9.0$  in increasing value as we go left to
right.  We see that there is a huge magnification of error in $\tildet1$ for a
corresponding small error in $m_t$ when the top mass is close to its fixed
point value.

\vskip 10pt

\noindent{\bf Fig. 2} Exhibition of the effect of the Landau pole on the
tachyonic limit for the case $m_0$=600 GeV,$m_{gluino}$ =300GeV and $\mu<0$.
The top curve gives the position of the Landau pole as a function of
$tan\beta$. The middle curve gives,  for the  case when $A_t=0.5 m_0$, the
value of $m_t$ as  a function of $tan\beta$ so that for all $m_t$ values above
this value $\tildet1$ is always tachyonic. The bottom curve is the same as
 the middle curve except that  $A_t=-0.5 m_0$.

\noindent{\bf Fig. 3}  The solutions for $m^2_{{\tilde t}_1}=0$ on the
$\mgluino$-$\m0$ plane for $\alphagut=24.11$, $\mgut=10^{16.187}$ GeV, and
$\mtop=170$~GeV and $\tan\beta=5.0$. This figure shows the branch for $\at>0$.
The light stop is tachyonic above and left to each curve for a fixed $\at$. All
masses are in GeV.

\noindent{\bf Fig. 4}  The solutions for $m^2_{{\tilde t}_1}=0$ on the
$\mgluino$-$\m0$ plane for $\alphagut=24.11$, $\mgut=10^{16.187}$ GeV, and
$\mtop=170$~GeV and $\tan\beta=5.0$. This figure shows the branch for $\at<0$.
The light stop is tachyonic above and right to each curve for a fixed $\at$.
All masses are in GeV.

\eject

\noindent{\bf Table 1  }
\vskip 10pt
\vbox{\offinterlineskip
      \hrule
      \halign{&\vrule#&
      \strut\hfil#\hfil\hfil\cr
    height2pt&\omit&&\omit&&\omit&\cr
     &\omit&&~~One Loop~~&&~~Two Loop ~~&\cr
     \noalign{\hrule}
    height2pt&\omit&&\omit&&\omit&\cr
     &~~$\alphagut^{-1}=25.7$~~&&\omit&&\omit&\cr
    height2pt&\omit&&\omit&&\omit&\cr
     &$\alpha_1$&&0.0166359&&0.0165156&\cr
    height2pt&\omit&&\omit&&\omit&&\omit&\cr
     &$\alpha_2$&&0.0323481&&0.031735 &\cr
    height2pt&\omit&&\omit&&\omit&\cr
     &$\alpha_3$&&0.0994&&\omit& \cr
     \noalign{\hrule}
    height2pt&\omit&&\omit&&\omit&\cr
     &$\alphagut^{-1}=24.5$&&\omit&&\omit& \cr
    height2pt&\omit&&\omit&&\omit&\cr
     &$\alpha_1$&&0.0169748&&0.0168391 &\cr
    height2pt&\omit&&\omit&&\omit&\cr
     &$\alpha_2$&&0.0336544&&0.0329366 &\cr
    height2pt&\omit&&\omit&&\omit&\cr
     &$\alpha_3$&&0.1129&&\omit&\cr
     \noalign{\hrule}
    height2pt&\omit&&\omit&&\omit&\cr
     &$\alphagut^{-1}=24.1$&&\omit&&\omit&\cr
    height2pt&\omit&&\omit&&\omit&\cr
     &$\alpha_1$&&0.0170909&&0.0169493 &\cr
    height2pt&\omit&&\omit&&\omit&\cr
     &$\alpha_2$&&0.0341137&&0.0333552 &\cr
    height2pt&\omit&&\omit&&\omit&\cr
     &$\alpha_3$&&0.1182&&\omit&\cr
     \noalign{\hrule}
    height2pt&\omit&&\omit&&\omit&\cr
     &$\alphagut^{-1}=24.0$&&\omit&&\omit&\cr
    height2pt&\omit&&\omit&&\omit&\cr
     &$\alpha_1$&&0.0171201&&0.016977  &\cr
    height2pt&\omit&&\omit&&\omit&\cr
     &$\alpha_2$&&0.0342304&&0.0334613 &\cr
    height2pt&\omit&&\omit&&\omit&\cr
     &$\alpha_3$&&0.1196&&\omit&\cr
    height2pt&\omit&&\omit&&\omit&\cr}\hrule}

\vskip 10pt
\noindent{\bf Table 2}
\vskip 10pt
\vbox{\offinterlineskip
      \hrule
      \halign{&\vrule#&
      \strut\hfil#\hfil\hfil\cr
    height2pt&&\omit&&\omit&&\omit&\cr
     &\omit&&~~One Loop~~&&~~Two Loop~~&\cr
     \noalign{\hrule}
    height2pt&\omit&&\omit&&\omit&\cr
      &~~$\alphagut^{-1}=24.11$~~&&\omit&&\omit&\cr
    height2pt&\omit&&\omit&&\omit&\cr
      &$E$&&12.985&&13.290&\cr
    height2pt&\omit&&\omit&&\omit&\cr
      &$F$&&266.48&&275.97&\cr
    height2pt&\omit&&\omit&&\omit&\cr
      &$F/E$&&20.523&&20.766&\cr
     \noalign{\hrule}
    height2pt&\omit&&\omit&&\omit&\cr
      &$\alphagut^{-1}=24.5$&&\omit&&\omit&\cr
    height2pt&\omit&&\omit&&\omit&\cr
      &$E$&&11.938&&12.536&\cr
    height2pt&\omit&&\omit&&\omit&\cr
      &$F$&&252.61&&266.78&\cr
    height2pt&\omit&&\omit&&\omit&\cr
      &$F/E$&&21.160&&21.288&\cr
     \noalign{\hrule}
    height2pt&\omit&&\omit&&\omit&\cr
      &$\alphagut^{-1}=25.0$&&\omit&&\omit&\cr
    height2pt&\omit&&\omit&&\omit&\cr
      &$E$&&11.101&&12.536&\cr
    height2pt&\omit&&\omit&&\omit&\cr
      &$F$&&242.46&&256.15&\cr
    height2pt&\omit&&\omit&&\omit&\cr
      &$F/E$&&21.842&&21.919&\cr
    height2pt&\omit&&\omit&&\omit&\cr}\hrule}

\vfill\eject\bye